\documentclass{aa}

\usepackage{graphicx}
\usepackage{verbatim}
\usepackage{color}
\usepackage{natbib}
\usepackage{shortbold}
\usepackage{booktabs}
\usepackage{tabularx}
\usepackage{amsmath}
\usepackage{array}
\usepackage{multirow}
\usepackage{makecell}
\usepackage[normalem]{ulem}
\usepackage{xspace}
\usepackage{xcolor}
\usepackage[varg]{txfonts}
\usepackage{colortbl,dcolumn}

\usepackage[bookmarks = true,colorlinks = true,linkcolor = red,urlcolor = blue,citecolor = blue,pdftex,bookmarks = true,linktocpage = true,hyperindex = true,breaklinks]{hyperref}

\def\zz#1{%
\ifdim#1pt>89pt\cellcolor{green}\else
\ifdim#1pt>79pt\cellcolor{yellow}\else
\cellcolor{red}\fi\fi
#1}

\bibpunct{(}{)}{;}{a}{}{,}

\begin{document}

\title{Multi-Channel Auto-Calibration for the Atmospheric Imaging Assembly using Machine Learning}

\author{Luiz F. G. dos Santos\inst{1,2}
        \and
        Souvik Bose\inst{3,4}
        \and
        Valentina Salvatelli\inst{5,6}
        \and
        Brad Neuberg\inst{5,6}
        \and
        Mark C. M. Cheung\inst{7}
        \and
        Miho Janvier\inst{8}
        \and
        Meng Jin\inst{6,7}
        \and
        Yarin Gal\inst{9}
        \and
        Paul Boerner\inst{7}
        \and
        At{\i}l{\i}m G\"{u}ne\c{s} Bayd{\rlap{\.}\i}n\inst{10,11}
        }

\institute{Heliophysics Science Division, NASA, Goddard Space Flight Center, Greenbelt, MD 20771, USA
        \and
        The Catholic University of America, Washington, DC 20064, USA % 620 Michigan Avenue NE,
        \and
        Rosseland Center for Solar Physics, University of Oslo,P.O. Box 1029 Blindern, NO-0315 Oslo, Norway
        \and
        Institute of Theoretical Astrophysics, University of Oslo,P.O. Box 1029 Blindern, NO-0315 Oslo, Norway
        \and
        Frontier Development Lab, Mountain View, CA 94043, USA
        \and
        SETI Institute, Mountain View, CA 94043, USA
        \and
        Lockheed Martin Solar \& Astrophysics Laboratory (LMSAL), Palo Alto, CA 94304, USA
        \and
        Universit\'e Paris-Saclay, CNRS, Institut d'astrophysique spatiale, Orsay, France
        \and
        OATML, Department of Computer Science, University of Oxford, Oxford OX1 3QD, UK
        \and
        Department of Computer Science, University of Oxford, Oxford OX1 3QD, UK
        \and
        Department of Engineering Science, University of Oxford, Oxford OX1 3PJ, UK}

\abstract
{Solar activity plays a quintessential role in influencing the interplanetary medium and space-weather around Earth. Remote sensing instruments on-board heliophysics space missions provide a pool of information about the Sun’s activity, via the measurement of its magnetic field and the emission of light from the multi-layered, multi-thermal, and dynamic solar atmosphere. Extreme UV (EUV) wavelength observations from space help in understanding the subtleties of the outer layers of the Sun, namely the chromosphere and the corona. Unfortunately, such instruments, like the Atmospheric Imaging Assembly (AIA) on-board NASA’s Solar Dynamics Observatory (SDO), suffer from time-dependent degradation that reduces their sensitivity. Current state-of-the-art calibration techniques rely on sounding rocket flights to maintain absolute calibration, which are infrequent, complex, and limited to a single vantage point.}
{We aim to develop a novel method based on machine learning (ML) that exploits spatial patterns on the solar surface across multi-wavelength observations to auto-calibrate the instrument degradation.}
{We establish two convolutional neural network (CNN) architectures that take either single-channel or multi-channel input and train the models using the SDOML dataset. The dataset is further augmented by randomly degrading images at each epoch with the training dataset spanning non-overlapping months with the test dataset. We also develop a non-ML baseline model to assess the gain of the CNN models. With the best trained models, we reconstruct the AIA multi-channel degradation curves of 2010--2020 and compare them with the sounding-rocket based degradation curves.}
{Our results indicate that the CNN-based models significantly outperform the non-ML baseline model in calibrating instrument degradation. Moreover, multi-channel CNN outperforms the single-channel CNN, which suggests the importance of cross-channel relations between different EUV channels for recovering the degradation profiles. The CNN-based models reproduce the degradation corrections derived from the sounding rocket cross-calibration measurements within the experimental measurement uncertainty, indicating that it performs equally well when compared with the current techniques.}
{Our approach establishes the framework for a novel technique based on CNNs to calibrate EUV instruments. We envision that this technique can be adapted to other imaging or spectral instruments operating at other wavelengths.}

\keywords{Sun: activity, UV radiation, and general - Techniques: image processing - Methods: data analysis - telescopes}

\date{Received 2 December 2020 / Accepted 29 January 2021} 

 \maketitle

\section{Introduction}
\label{Section:intro}
Solar activity plays a significant role in influencing the interplanetary medium and space weather around Earth and all the other planets of the solar system \citep{Schwenn2006}. Remote-sensing instruments on-board heliophysics missions can provide a wealth of information on the Sun’s activity, primarily via capturing the emission of light from the multi-layered solar atmosphere, thereby leading to the inference of various physical quantities such as magnetic fields, plasma velocities, temperature and emission measure, to name a few.
%and the measurement of magnetic and velocity fields.

\begin{figure*}
    \includegraphics[width=\textwidth]{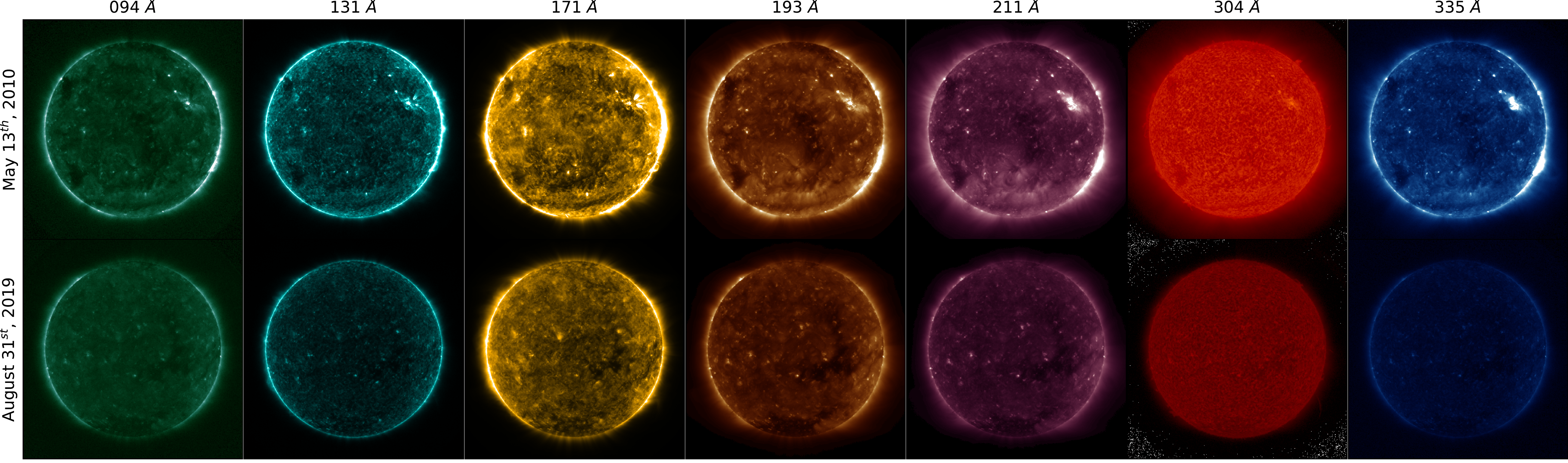}
    \caption{Set of images to exemplify how degradation affects the AIA channels. The two sets are composed of seven images from different EUV channels. From left to right: AIA $94$~\AA, AIA $131$~\AA, AIA $304$~\AA, AIA $335$~\AA, AIA $171$~\AA, AIA $193$~\AA, and AIA $211$~\AA. The top row corresponds to images from May $13^{th}$, $2010$ and the bottom row shows images from August $31^{st}$, $2019$, with no degradation correction. The $304$~\AA~channel images are in log-scale due the severe degradation.}
    \label{fig:autocalibrate_model_problem}
\end{figure*}

NASA currently manages the Heliophysics System Observatory (HSO), which consists of a group of satellites that constantly monitor the Sun, its extended atmosphere, space environments around Earth and other planets of the solar system \citep{HSO}. One of the flagship missions of HSO is the Solar Dynamics Observatory \citep[SDO, ][]{SDO_primary}. Launched in $2010$, SDO has been instrumental in monitoring the Sun's activity and providing a high volume of valuable scientific data every day with a high temporal and spatial resolution. It has three instruments onboard: the Atmospheric Imaging Assembly \citep[AIA,][]{AIA}, which records high spatial and temporal resolution images of the Sun in the ultraviolet (UV) and extreme UV (EUV); the Helioseismic \& Magnetic Imager \citep[HMI,][]{HMI}, that provides maps of the photospheric magnetic field, solar surface velocity, and continuum filtergrams; and the EUV Variability Experiment \citep[EVE,][]{EVE}, which measures the solar EUV spectral irradiance.

Over the past decade, SDO has played a central role in advancing our understanding of the fundamental plasma processes governing the Sun and space weather. This success can mainly be attributed to its open-data policy and a consistent high data-rate of approximately two terabytes of scientific data per day. The large volume of data accumulated over the past decade (over 12 petabytes) provides a fertile ground to develop and apply novel machine learning (ML) based data processing methods. Recent studies, such as, predicting solar flares from HMI vector magnetic fields \citep{Bobra_2015}, creating high-fidelity virtual observations of the solar corona \citep[\citealt{salvatelli2019} \&][]{Cheung2019}, forecasting far side magnetograms from the Solar Terrestrial Relations Observatory \citep[STEREO, ][]{Kaiser_2008} EUV images \citep{Kim_NatAs_2019}, super-resolution of magnetograms \citep{jungbluth-2019-super}, and mapping EUV images from AIA to spectral irradiance measurements \citep{Szenicereaaw6548}, have demonstrated the immense potential of ML applications in solar and heliophysics. In this paper, we leverage the availability of such high-quality continuous observations from SDO and apply ML techniques to address the instrument calibration problem.

One of the crucial issues that limit the diagnostic capabilities of the SDO-AIA mission is the degradation of sensitivity over time. Sample images from the seven AIA EUV channels in Fig.~\ref{fig:autocalibrate_model_problem} show an example of such a deterioration. The top row shows the images observed during the early days of the mission, from 13 May 2010, and the bottom row shows the corresponding images observed more recently on 31 August 2019, scaled within the same intensity range. It is clear that the images in the bottom row appear to be significantly dimmer compared to their top row counterparts. In some channels, especially $304$~\AA\ and $335$~\AA\, the effect is pronounced. 

The dimming effect observed among the channels is due to the temporal degradation of that EUV instruments in space that is also known to diminish the overall instrument sensitivity with time~\citep[e.g.,][]{BenMoussa_etal_2013}. The possible causes include either the out-gassing of organic materials in the telescope structure, which may deposit on the optical elements \citep{Jiao_2019}, or the decrease in detector sensitivity due to exposure to EUV radiation from the Sun. 

In general, first-principle models predicting the sensitivity degradation as functions of time and wavelength are not sufficiently well-constrained for maintaining the scientific calibration of such instruments. To circumvent this problem, instrument scientists have traditionally relied on empirical techniques, such as considering sources with known fluxes, the so-called "standard candles." However, no standard candles exist on the solar atmosphere at these wavelengths since the solar corona is continuously driven and structured by evolving magnetic fields, which caused localized and intermittent heating. This causes even the quiet Sun brightness in the EUV channels to vary significantly depending on the configuration of the small-scale magnetic fields \citep[][and the references therein]{2015A&A...581A..51S}. On the one hand, the Sun may not be bright enough to appear in the hotter EUV channels such as AIA 94~\AA\. On the other hand, active regions (ARs) have EUV fluxes that can vary by several orders of magnitude depending on whether it is in an emerging, flaring, or decaying state. Moreover, the brightness depends on the complexity of the AR's magnetic field \citep{2015LRSP...12....1V}. Finally, ARs in the solar corona can evolve on time scales ranging from a few minutes to several hours, leading to obvious difficulties in obtaining a standard flux for the purpose of calibration. 

Current state-of-the-art methods to compensate for this degradation rely on cross-calibration between AIA and EVE instruments. The calibrated measurement of the full-disk solar spectral irradiance from EVE is passed through the AIA wavelength (filter) response function to predict the integrated AIA signal over the full field of view. Later, the predicted band irradiance is compared with the actual AIA observations \citep{Boerner2013}. The absolute calibration of SDO-EVE is maintained through periodic sounding rocket experiments \citep{EVE_rocket} that use a near-replica of the instrument on-board SDO to gather a calibrated observation spanning the short interval of the suborbital flight (lasting a few minutes). A comparison of the sounding rocket observation with the satellite instrument observation provides an updated calibration, revealing long-term trends in the sensitivities of EVE and, thus, of AIA. 

Sounding rockets are undoubtedly crucial; however, the sparse temporal coverage (there are flights roughly every two years) and the complexities of inter-calibration are also potential sources of uncertainty in the inter-instrument calibration. Moreover, the inter-calibration analysis has long latencies, of months and sometimes years, between the flights and when the calibration can be updated, due to data analysis of the obtained data during the flight; also this kind of calibrations are limited to observations from Earth, and thus cannot easily be used to calibrate missions in deep space (e.g., STEREO).

In this paper, we focus on automating the correction of the sensitivity degradation of different AIA wavebands by using AIA information exclusively and adopting a deep neural network \citep[DNN, ][]{goodfellow2016deep} approach, which exploits the spatial patterns and cross-spectral correlations among the observed solar features in multi-wavelength observations of AIA. We compare our approach with a non-ML method motivated by solar physics heuristics, which we call the baseline model. We evaluate the predicted degradation curves with the ones obtained through the sounding rocket cross-calibration described above. To the best of our knowledge, this is the first attempt to develop a calibration method of this kind.\footnote{We presented an early-stage result of this work as an extended abstract at the NeurIPS workshop on ML and Physical Sciences 2019 (which has no formal proceedings) \citep[NeurIPS 2019, ][]{neuberg2019} where we described some preliminary results in this direction. In this paper, we extend the abstract with full analyses and discussion of several important issues, such as the performance on the real degradation curve and the limitations of the presented models, that are both crucial to evaluate the applicability of this ML-based technique.} We believe that the approach developed in this work could potentially remove a major impediment to developing future HSO missions that can deliver solar observations from different vantage points beyond Earth's orbit.
 
The paper is structured as follows: in Section \ref{Section:data}, we present and describe our dataset. In Section \ref{section:methodology} we illustrate the technique and how it has been developed. Namely, in \S~\ref{section:formulation} we state the hypothesis and propose a formulation of the problem, in \S~\ref{section:convolutional} we present the CNN models, in \S~\ref{Section:Analysis} we describe the training process and the evaluation, in \S~\ref{section:inter_channel} we probe the multi-channel relationship and in  \S~\ref{section:model-benchmark-understanding} we reconstruct the temporal degradation curve. Furthermore, in Section \ref{section:baseline} we present the baseline, followed by Section \ref{Section:Results} where we present and discuss the results. The concluding remarks are in Section \ref{section:summary}.
 
\section{Data description and pre-processing}
\label{Section:data}

We use for this study the pre-processed SDO-AIA dataset from \citet[][hereafter referred to as SDOML]{SDOML}. This dataset is ML-ready to be used for any kind of application related to the AIA and HMI data, and it consists of a subset of the original SDO data ranging from $2010$ to $2018$. It comprises the $7$~EUV channels, $2$~UV channels from AIA, and vector magnetograms from HMI. The data from the two SDO instruments are temporally aligned, with cadences of $6$ minutes for AIA (instead of the original $12$ seconds) and EVE and $12$ minutes for HMI. The full-disk images are downsampled from $4096 \times 4096$ to $512 \times 512$ pixels and have an identical spatial sampling of $\thicksim$ $4\farcs8$ per pixel. 

In SDOML, the AIA images have been compensated for the exposure time and corrected for instrumental degradation over time using piecewise-linear fits to the V8 corrections released by the AIA team in November 2017.\footnote{Available at \url{https://aiapy.readthedocs.io/en/stable/generated/gallery/instrument\_degradation.html\#sphx-glr-generated-gallery-instrument-degradation-py}} These corrections are based on cross-calibration with SDO-EVE, where the EVE calibration is maintained by periodic sounding rocket under flights (including, in the case of the V8 corrections, a flight on 1 June 2016). Consequently, the resulting dataset offers images where changes in pixel brightness are directly related to the state of the Sun rather than instrument performance.

In this paper, we applied a few additional pre-processing steps. First, we downsampled the SDOML dataset to $256\times256$ pixels from $512\times512$ pixels. We established that $256\times256$ is a sufficient resolution for the predictive task of interest (inference of a single coefficient), and the reduced size enabled quicker processing and more efficient use of the computational resources. Secondly, we masked the off-limb signal ($r>R_\odot$) to avoid possible contamination due to the telescope vignetting. Finally, we re-scaled the brightness intensity of each AIA channel by dividing the image intensity by a channel-wise constant factor. These factors represent the approximate average AIA data counts in each channel and across the period from 2011 to 2018 \citep[derived from][]{SDOML}, and this re-scaling is implemented to set the mean pixel values close to unity in order to improve the numerical stability and the training convergence of the CNN. Data normalization such as this is standard practice in NNs \citep{goodfellow2016deep}. The specific values for each channel are reported in Appendix~\ref{section:appendix_average}.

\section{Methodology}
\label{section:methodology}
\subsection{Formulation of the problem}
\label{section:formulation}
\begin{figure*}
    \includegraphics[width=\textwidth]{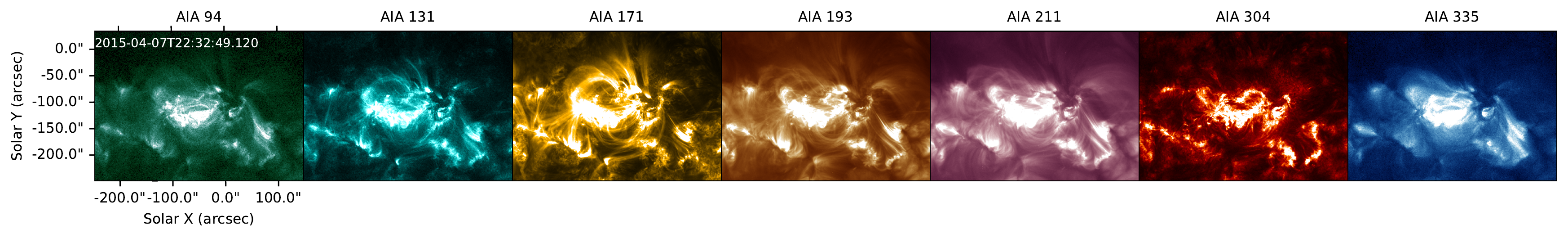}
    \caption{A co-located set of images of the seven EUV channels of AIA to exemplify structures that are observed across different wavelengths. From left to right: AIA $94$~\AA, AIA $131$~\AA, AIA $304$~\AA, AIA $335$~\AA, AIA $171$~\AA, AIA $193$~\AA, and AIA $211$~\AA.}
    \label{fig:morphology}
\end{figure*}

It is known that some bright structures in the Sun are observed across different wavelengths. Figure \ref{fig:morphology} shows a good example from $07$ April $2015$ of a bright structure in the center of all seven EUV channels from AIA. Based on this cross-channel structure, we establish a hypothesis divided into two parts. First is that there is a relationship between the morphological features and the brightness of solar structures in a single channel (e.g., typically, dense and hot loops over ARs). The second is that such a relationship between the morphological features and the brightness of solar structures can be found across multiple channels of AIA. We hypothesize that both these relationships can be used to estimate the dimming factors and that a deep learning model can automatically learn these inter- and cross-channel patterns and exploit them for accurately predicting the dimming factor of each channel. 

To test our hypothesis we consider a vector $\vec{C} = \{C_i, i\in [1,...,n]\}$ of multi-channel, synchronous SDO/AIA images, where $C_i$ denotes the $i$-th channel image in the vector, and a vector $\vec{\alpha} = \{\alpha_i, i \in [1,...,n]\}$, where $\alpha_i$ is the dimming factor independently sampled from the continuous uniform distribution between [$0.01, 1.0$]. We choose an upper bound value of  $\alpha_i = 1$, since we only consider dimming of the images and not enhancements. Further we create a corresponding vector of dimmed images as $\vec{D} = \{\alpha_i C_i, i\in [1,...,n]\}$, where $\vec{D}$ is the corresponding dimmed vector. It is also to be noted that the dimming factors $\alpha_i$ are applied uniformly per channel and are not spatially dependent. The spatial dependence of the degradation is assumed to be accounted for by regularly updated flat-field corrections applied to AIA images. Our goal in this paper is to find a deep learning model $M: \vec{D} \rightarrow \vec{\alpha}$ that retrieves the vector of multi-channel dimming factors $\vec{\alpha}$ from the observed SDO-AIA vector $\vec{D}$. 

\subsection{Convolutional Neural Network Model}
\label{section:convolutional}
Deep learning is a very active sub-field of machine learning that focuses on specific models called deep neural networks (DNNs). A DNN is a composition of multiple layers of linear transformations and non-linear element-wise functions \citep{goodfellow2016deep}. One of the main advantages of deep learning is that it can learn from the data the best feature representation for a given task without the need to manually engineer such features. DNNs have produced state-of-the-art results in many complex tasks including object detection in images \citep{he2016deep}, speech recognition \citep{amodei2016deep} and synthesis \citep{oord2016wavenet}, translation between languages \citep{wu2016google}. A DNN expresses a differentiable function $F_{\vec\theta}: \mathcal{X} \to \mathcal{Y}$ that can be trained to perform complex non-linear transformations, by tuning parameters $\vec{\theta}$ using gradient-based optimization of a loss (also known as objective or error) function $L(\vec{\theta}) = \sum_i l(F_{\vec\theta}(\vec{x}_i), \vec{y}_i)$ for a given set of inputs and desired outputs $\{\vec{x}_i, \vec{y}_i\}$.

\begin{figure}
    \centering
        \includegraphics[width=0.5\textwidth]{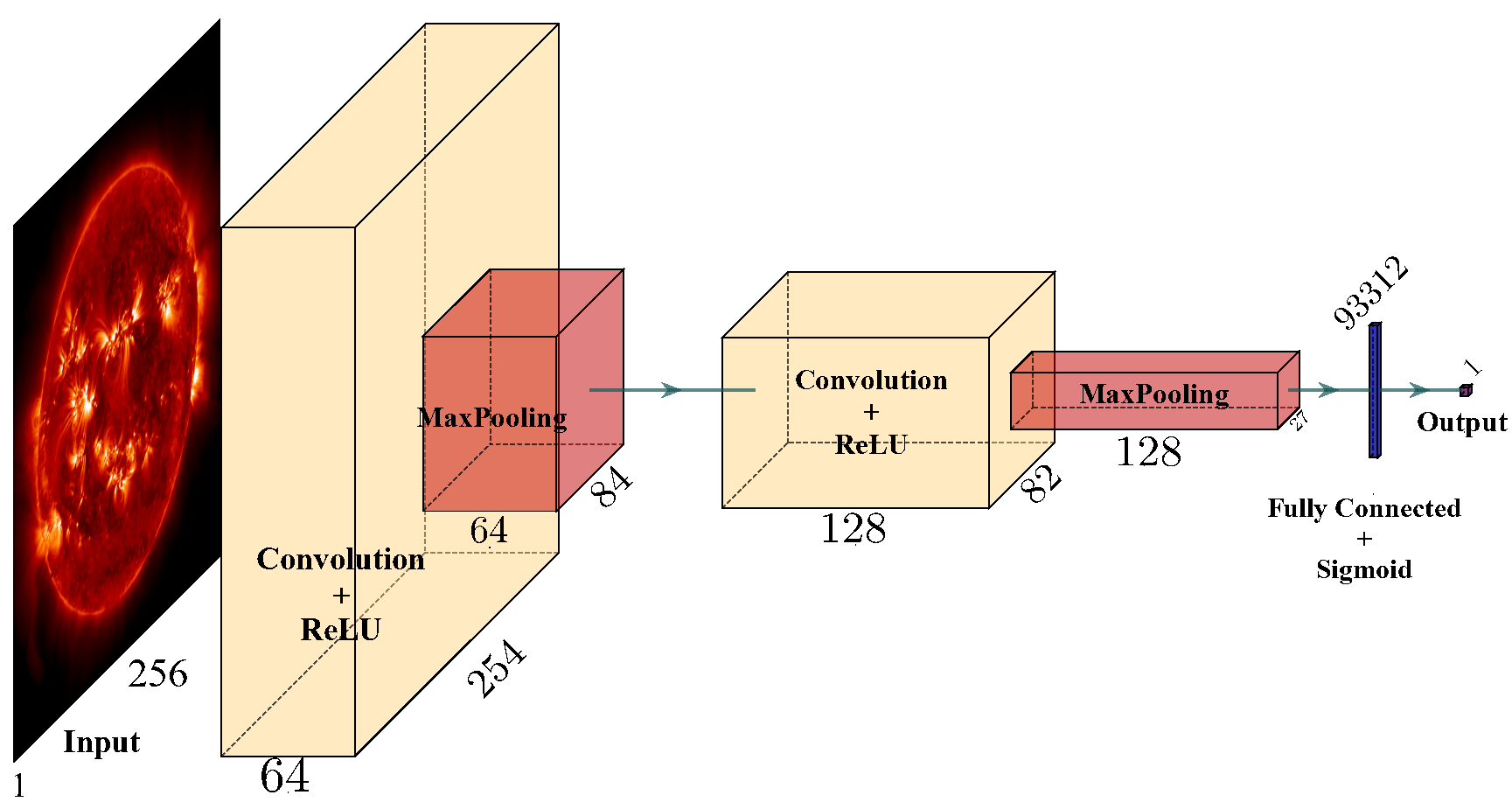}
        \includegraphics[width=0.5\textwidth]{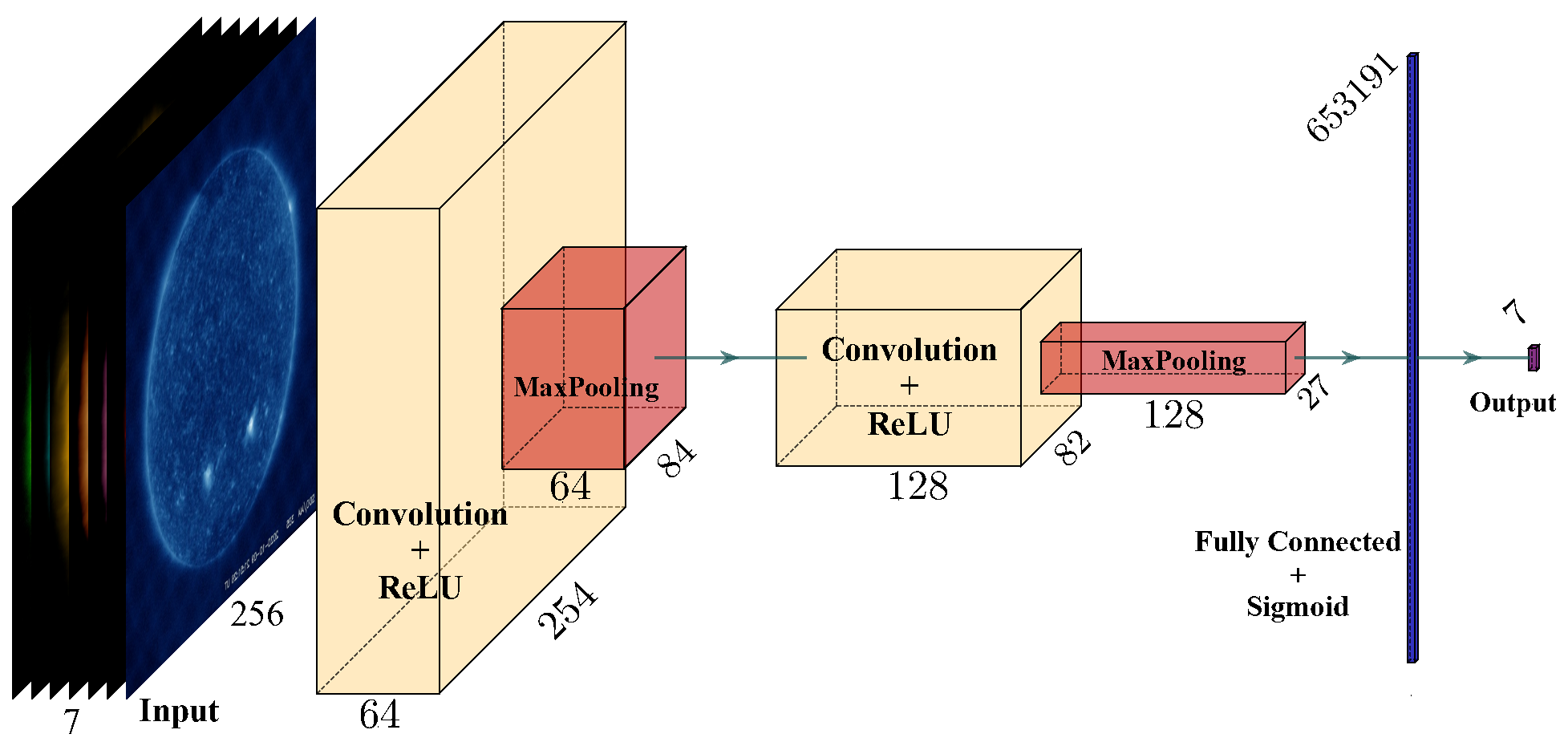}
        \caption{The CNN architectures used in this paper. At the top the single-channel architecture with a single wavelength input and composed of two blocks of a convolutional layer, ReLU activation function and max pooling layer, followed by a fully connected (FC) layer and a final sigmoid activation function. At the bottom the multi-channel architecture with a multi wavelength input and composed of two blocks of a convolutional layer, ReLU activation function and max pooling layer, followed by a fully connected (FC) layer and a final sigmoid activation function. Figures constructed with \cite{haris_iqbal_2018_2526396}}
        \label{fig:autocalibrate_CNN_arch}
\end{figure}

For the degradation problem summarized in Section~\ref{section:formulation}, we consider two CNN architectures \citep{lecun1995convolutional}. The first architecture does not exploit the spatial dependence across multi-channel AIA images, therefore ignoring any possible relationship that different AIA channels might have, and it is designed to explore only the relationship across different structures in a single channel. This architecture is a test for the first hypothesis in Section~\ref{section:formulation}. The second architecture is instead designed to exploit possible cross-channel relationships while training, and it tests our second hypothesis, that solar surface features appearing across the different channels will make a  multi-channel CNN architecture more effective than a single channel CNN that only exploit inter-channel structure correlations. The first model considers a single channel as input in the form of a tensor with shape $1\times256\times256$ and has a single degradation factor $\alpha$ as output. The second model takes in multiple AIA channel images simultaneously as an input with shape $n\times256\times256$ and output $n$ degradation factors $\vec{\alpha} = \{\alpha_i, i \in [1,...,n]\}$, where $n$ is the number of channels as indicated in Fig.~\ref{fig:autocalibrate_CNN_arch}. 

The single- and multi-channel architectures are described in (Fig.~\ref{fig:autocalibrate_CNN_arch}). They both consist of two blocks of a convolutional layer followed by ReLU (rectified linear unit) activation function \citep{Nair:2010:RLU:3104322.3104425} and a max pooling layer. These are followed by a fully connected (FC) layer and a final sigmoid activation function that is used to output the dimming factors. The first convolution block has 64 filters, while the second convolution block has 128. In both convolution layers, the kernel size is $3$, meaning the filters applied on the image are $3\times3$ pixels, and the stride is $1$, meaning that the kernel slides through the image 1 pixel per step. No padding is applied (i.e., no additional pixels are added at the border of the image to avoid a change in size). The resulting total learnable parameters (LP) are $167,809$ for the single-channel model and $731,143$ for the multi-channel model. The final configurations of the models' architectures were obtained through a grid search among different hyperparameters and layer configurations. More details of the architectures can be found in Appendix \ref{section:appendix_archtectures}. 

We use the open-source software library PyTorch \citep{paszke_2017} to implement the training and inference code for the CNN. The source code to produce this paper is publicly available at \cite{autocal_code} and \url{https://github.com/vale-salvatelli/sdo-autocal_pub}.

\subsection{Training Process}
\label{Section:Analysis}
The actual degraded factors $\alpha_i(t)$ (where $t$ is the time since the beginning of the SDO mission, and $i$ is the channel) trace a single trajectory in an $n$-dimensional space starting with $\alpha_i(t=0)=1$ $\forall$ $i\in[1,...,n]$ at the beginning of the mission. During training, we intentionally exclude this time-dependence from the model. This is done by ($1$) using the SDOML dataset, which has already been corrected for degradation effects, ($2$) not assuming any relation between $t$ and $\vec{\alpha}$ and not using $t$ as an input feature, and ($3$) temporally shuffling the data used for training. As presented in section \ref{section:formulation}, we degrade the each set of multi-channel images~$\vec{C}$ by a unique $\vec{\alpha} = \{\alpha_i, i \in [1,...,n]\}$. We then devised a strategy such that from one training epoch to the next, the same set of multi-channel images can be dimmed by a completely independent set of $\vec{\alpha}$ dimming factors. This is a data augmentation and regularization procedure that allows the model to generalize and perform well in recovering dimming factors over a wide range of solar conditions.

The training set comprises multi-channel images~$\vec{C}$ obtained during the months of January to July from $2010$ to $2013$ obtained every six hours, amounting to a total of $18,970$ images in $2,710$ timestamps.The model was trained using 64 samples per minibatch, and the training has been performed for $1,000$ epochs. We do not use the full dataset to calculate the gradient descent and propagate back to update the network's parameters/weights in the minibatch concept. Instead, we calculate the gradient descent and correct the weights as the model is still going through the data. This procedure allows lowering the computation cost while still obtaining a lower variance. As a consequence of our data augmentation strategy, after $1000$ epochs the model has been trained with $2,710,000$ unique sets of (input, output) pairs since we used a different set of $\vec{\alpha}$ each epoch. We used the Adam optimizer \citep{Optimizer} in our training with an initial learning rate of $0.001$ and the mean squared error (MSE) of the predicted degradation factor ($\alpha_P$), and the ground truth value ($\alpha_{GT}$) was used as the training objective (loss).

The test dataset, i.e., the sample of data used to provide an unbiased evaluation of a model fit on the training dataset, holds images obtained during the months of August to October between $2010$ and $2013$, again every six hours per day, totaling $9,422$ images over $1,346$ timestamps. The split by month between the training and test data has a two-fold objective: ($1$) it prevents the bias due to the variation in the solar cycle, thereby allowing the model to be deployed in future deep space missions forecasting $\alpha$ for future time steps, and ($2$) it ensures that the same image is never present in both the datasets (any two images adjacent in time will approximately be the same), leading to a more precise and a comprehensive evaluation metric.
 
 \subsection{Toy Model Formulation to Probe the Multi-Channel Relationship}

\label{section:inter_channel}

Using the described CNN model, we tested the hypothesis using a toy dataset, which is simpler than the SDOML dataset. We tested if the physical relationship between the morphology and brightness of solar structures (e.g., ARs, coronal holes) across multiple AIA channels would help the model prediction. For this purpose, we created artificial solar images, in which a $2$D Gaussian profile is used (Equation \ref{E-relationship}) to mimic the Sun as an idealized bright disk with some center-to-limb variation:

\begin{equation}
\label{E-relationship}
    C_i(x,y) = A_i \exp{(-[x^2+y^2]{\sigma^{-2}})},
\end{equation}

\noindent where $A$ is the amplitude centered at ($0,0$), characteristic width $\sigma$, and $x$ and $y$, are the coordinates at the image. $\sigma$ is sampled from a uniform distribution between $0$ and $1$. These images are not meant to be a realistic representation of the Sun. However, as formulated in Eq.~\ref{E-relationship}, they include two qualities we posit to be essential for allowing our auto-calibration approach to be effective. The first is the correlation of intensities across wavelength channels (i.e., ARs tend to be bright in multiple channels). The second is the existence of a relationship between the spatial morphology of EUV structures with their brightness. This toy dataset is designed so that we can independently test how the presence of (a) a relation between brightness $A_i$ and size $\sigma$, and  (b) a relation between $A_i$ for various channels; and the presence of both (a) and (b) influences performance. To evaluate this test, we will use the MSE loss and expect the presence of both (a) and (b) to minimize this loss.

The test result of the multi-channel model with artificial solar images is shown in Table~\ref{table:toy_problem_metrics}. We can see that when $A_0 \propto \sigma$ (linear relation between size and brightness) and $A_i = A_0^i$ (i.e., dependence across channels; here $i$ superscript denotes $A_0$ raised to the $i$-th power), the CNN solution delivered minimal MSE loss (top-left cell). Eliminating the inter-channel relationship (i.e., each $A_i$ was randomly chosen) or the relation between brightness $A_i$ and size $\sigma$, the performance suffered increasing the MSE loss. Ultimately, when both $A_i$ and $\sigma_i$ were randomly sampled for all channels, the model performed equivalently to randomly guessing/regressing (bottom-right cell) and having the greater loss of all tests. These experiments confirm our hypothesis and indicate that a multi-channel input solution will outperform a single-channel input model in the presence of relationships between the morphology of solar structures and their brightness across the channels. 

\begin{table}
\centering
\caption{The mean squared error (MSE) for all combinations proposed in Section \ref{section:inter_channel}. The top-left cell is for the scenario when there exists a cross-channel correlation and a relation between brightness and size of the artificial Sun. The top-right cell, has is the loss with a cross-channel correlation but not the relation between brightness and size. The bottom left cell has the loss when there is no cross-channel correlation, but it has a relation between brightness and size. The bottom right cell presents the loss when the parameters are freely chosen.}
\label{table:toy_problem_metrics}
\begin{tabular}{|cc|c|l|}
\hline
 &
   &
  \multicolumn{2}{c|}{\begin{tabular}[c]{@{}c@{}}Brightness and size\\ correlation\end{tabular}} \\ \cline{3-4} 
                       &    & Yes   & No    \\ \hline
\multicolumn{1}{|c|}{\multirow{2}{*}{\begin{tabular}[c]{@{}c@{}}Cross-channel\\ correlation\end{tabular}}} &
  Yes &
  0.017 &
  0.023 \\ \cline{2-4} 
\multicolumn{1}{|c|}{} & No & 0.027 & 0.065 \\ \hline
\end{tabular}
\end{table}
 
\subsection{Reconstruction of the Degradation Curve using the CNN Models}
\label{section:model-benchmark-understanding}
 
In order to evaluate the model in a different dataset from the one used in the training process, we use both single-channel and multi-channel CNN architectures to recover the instrumental degradation over the entire period of SDO (from $2010$ to $2020$). To produce the degradation curve for both CNN models, we use an equivalent dataset of the SDOML dataset but without compensating the images for degradation\footnote{The SDOML dataset not corrected for degradation overtime is available at \url{https://zenodo.org/record/4430801\#.X_xuPOlKhmE} } \citep{SDOML_degraded} and having data from 2010 to 2020. All other pre-processing steps, including masking the solar limb, re-scaling the intensity, etc., remain unchanged. The CNN's estimates of degradation are then compared to the degradation estimates obtained from cross-calibration with irradiance measurements, computed by the AIA team using the technique described in \citep{Boerner2013}.

The cross-calibration degradation curve relies on the daily ratio of the AIA observed signal to the AIA signal predicted by SDO-EVE measurements up through the end of EVE MEGS-A operations in May $2014$. From May $2014$ onwards, the ratio is computed using the FISM model \citep{Chamberlin2020} in place of the EVE spectra. FISM is tuned to SDO-EVE, so the degradation derived from FISM agrees with the degradation derived from EVE through $2014$. However, the uncertainty in the correction derived from FISM is greater than that derived from EVE observations, primarily due to the reduced spectral resolution and fidelity of FISM compared to SDO-EVE. While the EVE-to-AIA cross-calibration introduced errors of approximately $4\%$ (on top of the calibration uncertainty intrinsic to EVE itself), the FISM-to-AIA cross-calibration has errors as large as $25\%$. 

We examined both $V8$ and $V9$ of the cross-calibration degradation curve. The major change from $V8$ calibration (released in November $2017$, with linear extrapolations extending the observed trend after this date) to $V9$ (July $2020$) is based on the analysis of the EVE calibration sounding rocket flown on $18$ June $2018$. The analysis of this rocket flight resulted in an adjustment in the trend of all channels during the interval covered by the FISM model (from May $2014$ onwards), as well as a $20\%$ shift in the $171$~\AA\ channel normalization early in the mission. This changes become more clear when looking at Fig.~\ref{fig:degradation_curve} at Sec.\ref{Section:Results}. The uncertainty of the degradation correction during the period prior to May $2014$, and on the date of the most recent EVE rocket flight, is dominated by the $\sim10\%$ uncertainty of the EVE measurements themselves. For periods outside of this (particularly periods after the most recent rocket flight), the uncertainty is a combination of the rocket uncertainty and the errors in FISM in the AIA bands (approximately $25\%$).
 
Moreover, we obtain and briefly analyze the feature maps from the second max pooling layer from the multi-channel model. A feature map is simply the output of one mathematical filter applied to the input. Looking at the feature maps, we expand our understanding of the model operation. This process helps to shine light over the image processing and provides insight into the internal representations combining and transforming information from seven different EUV channels into the seven dimming factors.

\section{Baseline Model}
\label{section:baseline}

We compare our DNN approach to a baseline motivated by the assumption that the EUV intensity outside magnetically ARs, i.e. the quiet Sun, is invariant in time \citep[a similar approach is also considered for the in-flight calibration of some UV instruments, e.g.][]{Schule1998}. A similar assumption in measuring the instrument sensitivity of the Solar \& Heliospheric Observatory \citep[SOHO, ][]{soho} CDS was also adopted by \citet{2010A&A...518A..49D}, where they assumed that the irradiance variation in the EUV wavelengths is mainly due to the presence of ARs on the solar surface and the mean irradiance of the quiet Sun is essentially constant over the solar cycle. Though there are evidences of small-scale variations in the intensity of quiet Sun when observed in the transition region \citep{2015A&A...581A..51S}, their contribution is insignificant in comparison to their AR counterparts. We use this idea for our baseline model as described in this section.

\begin{figure}[h]
    \centering
        \includegraphics[height=3.3in]{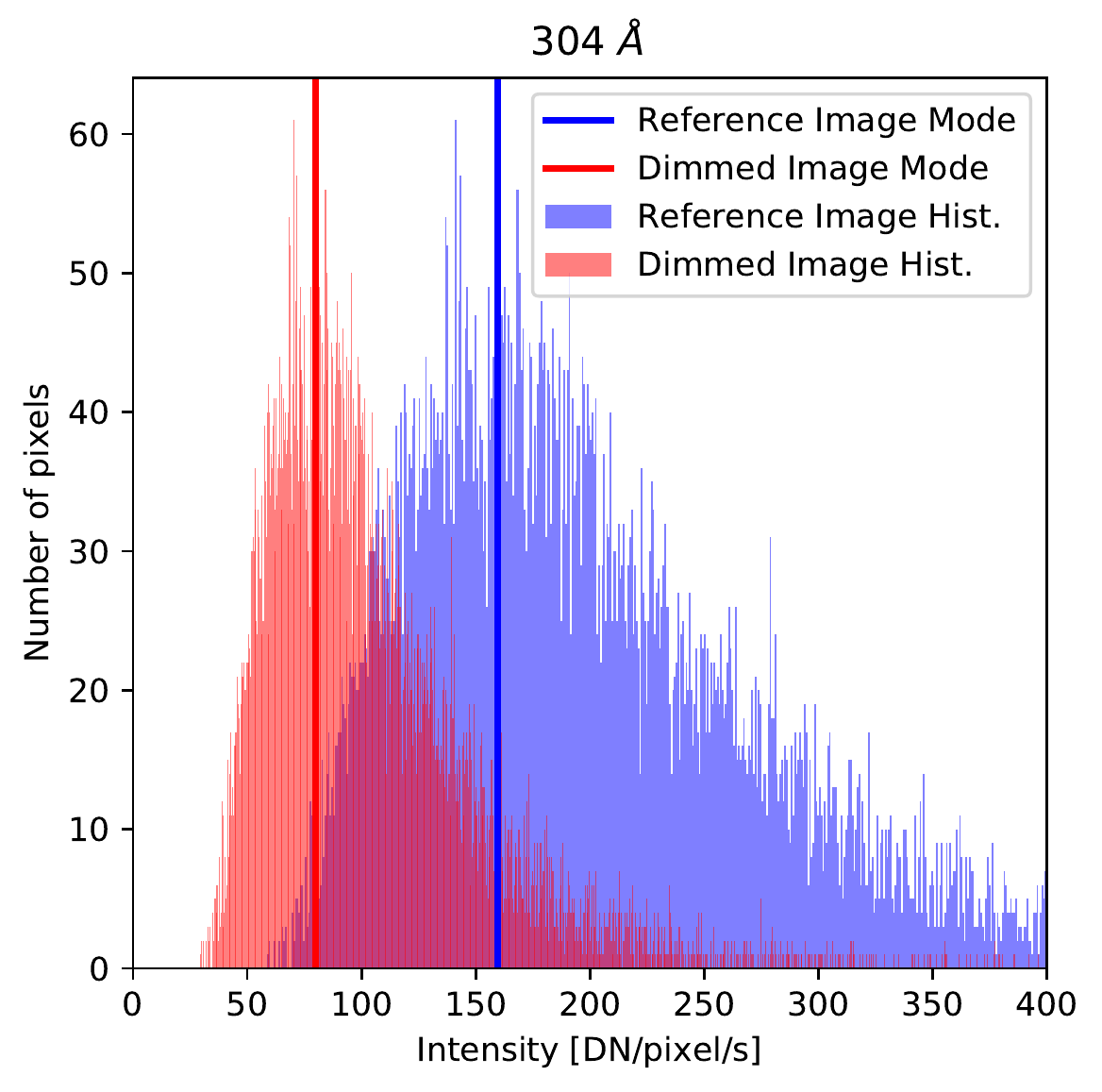}
        \caption{Histograms of the pixel values for $304$~\AA~  channel. In blue the histogram for the refence image and in red the histogram for the dimmed image. The y-axis is the number of pixels, and the x-axis is the pixel intensity [$DN/px/s$]. The modes are marked with blue and red line for the reference and dimmed images respectively.}
        \label{fig:baseline_histogram}
\end{figure}

It is important to remark that we use exactly the same data pre-processing and splitting approach as the one used for the neural network model described in Sect.~\ref{Section:Analysis}. From the processed dataset, a set of reference images per channel, ${\vec{C}_{\rm ref}}$, are selected at time $t=t_{\rm ref}$. Since the level of solar activity continuously evolves in time, we only select the regions of the Sun that correspond to low activity, as discussed in the preceding paragraph. Furthermore, the activity level is decided based on co-aligned (with AIA) magnetic field maps from HMI. To define these regions, we first make a square selection with a diagonal of length $2R_\odot$ centered at $R=0$ of the solar images so as to avoid LOS projection effects towards the limb. We then apply an absolute global threshold value of 5 Mx~cm$^{-2}$ on the co-aligned HMI LOS magnetic field maps corresponding to $t=t_{\rm ref}$, such that only those pixels that have B$_{\mathrm{LOS}}$ less than the threshold are extracted, resulting in a binary mask with 1 corresponding to the pixels of interest and 0 the rest. This minimum chosen value of the magnetic flux density is close to the noise level of the HMI\_720s magnetograms \citep{2012SoPh..279..295L,2018ApJ...862...35B}. Finally, we use this mask to extract the co-spatial quiet Sun (less active) pixels from each AIA channel and compute the respective 1D histograms of the intensity values as shown in Fig.~\ref{fig:baseline_histogram}. Now, based on the assumption that the intensity of the quiet Sun area does not change significantly over time (as discussed in the preceding section), we chose to artificially dim these regions by multiplying them with a constant random factor between 0 and 1. Naturally, values close to 0 will make the images progressively dimmer. The histograms for the dimmed and the original (undimmed) quiet Sun intensities for the AIA~304~\AA\ channel are shown in Fig.~\ref{fig:baseline_histogram}. The idea is to develop a non-machine learning approach that could be used to retrieve this dimming factor.

From Fig.~\ref{fig:baseline_histogram} we find that both the dimmed and undimmed 1D histograms have a skewed shape, with a dominant peak at lower intensities and extended tails at higher intensities. Such skewed distribution for the quiet Sun intensities has been reported by various studies in the past \citep[see][]{2015A&A...581A..51S}, where they have been modeled as either a sum of two Gaussians \citep{1976RSPTA.281..319R} or a single log-normal distribution \citep{1999ApJ...512..992G,2007A&A...468..695F}. Despite an increased number of free parameters in double Gaussian fitting, \citet{2000A&A...362..737P} showed that the observed quiet Sun intensity distribution could be fitted significantly better with a single log-normal distribution. The skewed representation, such as the one shown for the 304~\AA\ channel, was also observed for all the other EUV channels, indicating that the criterion for masking the quiet Sun pixels described here is justified.

We then compute the mode (most probable value) of both the dimmed and undimmed log-normal distributions and indicate them by  $I_{i,{\rm ref}}^{\rm mp}$ (where \textit{i} implies the AIA channel under consideration and \textit{mp} stands for the modal value for the undimmed images), and $I^{\rm mp}_{i}$ representing the modal intensity value for the corresponding images dimmed with a dimming factor (say $\alpha_i$). These are indicated by blue and red vertical lines in Fig.~\ref{fig:baseline_histogram}. Subsequently, the dimming factor is obtained by computing the ratio between the two most probable intensity values according to the following equation:

\begin{equation}
    \alpha_i := \frac{I^{\rm mp}_{i}}{I_{i, {\rm ref}}^{\rm mp}}
\end{equation}

Since both distributions are essentially similar except for the dimming factor, we suggest that such a ratio is efficient enough to retrieve $\alpha_i$ reliably forming a baseline against which the neural network models are compared. The efficiency of the baseline in recovering the dimming factor is then evaluated according to the success rate metric and the results for all channels are tabulated in Table~\ref{tab:autocalibrate_final_results}. 

\section{Results and Discussions}
 \label{Section:Results}

\subsection{Comparing the performances of the baseline model with different CNN architectures}
The results of the learning algorithm are binarized using five different thresholds: the absolute value of $0.05$ and relative values of $5\%$, $10\%$, $15\%$, and $20\%$. If the absolute difference between the predicted degradation factor ($\alpha_P$) and the ground truth degradation factor ($\alpha_{GT}$) is smaller than the threshold, it is considered a success $\alpha_P$; otherwise, it is not a success. We then evaluate the binarized results by using the success rate, which is the ratio of success $\alpha_P$ and the total amount of $\alpha_{P}$. We chose different success rate thresholds to gauge the model, all of which are smaller than the uncertainty of the AIA calibration \citep[estimated as $28\%$ by ][]{AIA_calib_paper}.

The baseline, single-channel, and multi-channel model results are summarized in Table~\ref{tab:autocalibrate_final_results}. The different colors are for different success rates: green is for success rates greater than $90\%$, yellow for success rate between $80\%$ and $90\%$, and red is for success rate lower than $80\%$.

\begin{table*}
  \centering
  \caption{Results of the baseline and CNN models applied to all the EUV AIA channels. The Table is divided into three sections: Baseline, Single-Channel, and Multi-Channel model. From the left, the channel number, the success rates for the baseline, the success rates for the single-channel CNN model, and the success rates for the multi-channel CNN model. Each model performance is considered at different tolerance levels. At the bottom, the mean of the success rate across all the channels. The color green is for success rates greater than $90\%$, yellow for success rate between $80\%$ and $90\%$, and red is for success rate lower than $80\%$.}
  \label{tab:autocalibrate_final_results}
  \centering
  \begin{tabular}{cccccccccccccccc}
    \toprule
    \multirow{2}{*}{Channel} & \multicolumn{5}{c}{\parbox{5cm}{\centering Baseline}} & \multicolumn{5}{c}{\parbox{5cm}{\centering Single-Channel Model}} &  \multicolumn{5}{c}{\parbox{5cm}{\centering Multi-Channel Model}}  \\
    \cmidrule(lr){2-6}\cmidrule(lr){7-11}\cmidrule(lr){12-16}
     & 0.05 & 5\% & 10\% & 15\% & 20\% & 0.05 & 5\% & 10\% & 15\% & 20\% & 0.05 & 5\% & 10\% & 15\% & 20\%\\
     \midrule
     94~\AA  & \zz {32}\% & \zz{08}\% & \zz{18}\% & \zz{28}\% & \zz {40}\% & \zz {70}\% & \zz {37}\% & \zz {61}\% & \zz {78}\% & \zz {87}\% & \zz {82}\% & \zz {48}\% & \zz {73}\% & \zz {85}\% & \zz {92}\% \\
     131~\AA & \zz {76}\% & \zz{50}\% & \zz{73}\% & \zz{86}\% & \zz {96}\% & \zz {94}\% & \zz {72}\% & \zz {92}\% & \zz {98}\% & \zz {99}\% & \zz {99}\% & \zz {76}\% & \zz {94}\% & \zz {97}\% & \zz {99}\% \\
     171~\AA & \zz {58}\% & \zz{27}\% & \zz{48}\% & \zz{66}\% & \zz {85}\% & \zz {93}\% & \zz {70}\% & \zz {93}\% & \zz {97}\% & \zz {99}\% & \zz {84}\% & \zz {48}\% & \zz {72}\% & \zz {86}\% & \zz {93}\% \\
     193~\AA & \zz {38}\% & \zz{13}\% & \zz{27}\% & \zz{44}\% & \zz {53}\% & \zz {73}\% & \zz {41}\% & \zz {69}\% & \zz {85}\% & \zz {93}\% & \zz {90}\% & \zz {59}\% & \zz {85}\% & \zz {94}\% & \zz {98}\% \\
     211~\AA & \zz {31}\% & \zz{11}\% & \zz{21}\% & \zz{29}\% & \zz {39}\% & \zz {63}\% & \zz {30}\% & \zz {53}\% & \zz {71}\% & \zz {84}\% & \zz {76}\% & \zz {41}\% & \zz {68}\% & \zz {82}\% & \zz {92}\% \\
     304~\AA & \zz {86}\% & \zz{66}\% & \zz{89}\% & \zz{95}\% & \zz{100}\% & \zz {90}\% & \zz {65}\% & \zz {89}\% & \zz {97}\% & \zz {99}\% & \zz {94}\% & \zz {62}\% & \zz {86}\% & \zz {93}\% & \zz {96}\% \\
     335~\AA & \zz {38}\% & \zz{13}\% & \zz{29}\% & \zz{42}\% & \zz {51}\% & \zz {62}\% & \zz {31}\% & \zz {54}\% & \zz {69}\% & \zz {80}\% & \zz {73}\% & \zz {39}\% & \zz {65}\% & \zz {82}\% & \zz {91}\% \\
     \textbf{Mean} & \textbf{\zz {51}\%} & \textbf{\zz {27}\%} & \textbf{\zz {43}\%} & \textbf{\zz {56}\%} & \textbf{\zz {66}\%} & \textbf{\zz {78}\%} & \textbf{\zz {50}\%} & \textbf{\zz {73}\%} & \textbf{\zz {85}\%} & \textbf{\zz {92}\%} & \textbf{\zz {85}\%} & \textbf{\zz {53}\%} & \textbf{\zz {77}\%} & \textbf{\zz {89}\%} & \textbf{\zz {94}\%} \\
     \bottomrule
  \end{tabular}
\end{table*}

A detailed look at Table~\ref{tab:autocalibrate_final_results} reveals that for an absolute tolerance value of $0.05$, the best results for the baseline are $86\%$ ($304$~\AA) and $76\%$ ($131$~\AA), and a mean success rate of $\sim51\%$ across all channels. As we increase the relative tolerance levels, the mean success rate increases from $27\%$ (for $5\%$ relative tolerance) to $66\%$ (with $20\%$ relative tolerance) and with a $39\%$ success rate in the worst-performing channel ($211$~\AA).

Investigating the performance of the CNN architecture with a single input channel and an absolute tolerance level of $0.05$, we find that this model performed significantly better than our baseline with much higher values of the metric for all the channels. The most significant improvement was shown by the $94$~\AA~ channel with an increase from $32\%$ in the baseline model to about $70\%$ in the single input CNN model, with an absolute tolerance of $0.05$. The average success rate bumped from $51\%$ in the baseline to $78\%$ in the single-channel model. The worst metric for the single-channel CNN architecture was recorded by the $211$~\AA\ channel, with a success rate of just $63\%$, which is still significantly better than its baseline counterpart ($31\%$). Furthermore, with a relative tolerance value of $15\%$, we find that the mean success rate is $85\%$ for the single-channel model, which increases to more than $90\%$ for a $20\%$ tolerance level. This is a promising result considering the fact that the error associated with the current state-of-the-art calibration techniques (sounding rockets) is $\sim25\%$.  

Finally, we report the results from the multi-channel CNN architecture in the last section of Table~\ref{tab:autocalibrate_final_results}. As expected, the performance, in this case, is the best of the models, with significant improvements for almost all the EUV channels. Clearly, the success rates belonging to the red category are much lesser compared to the former models implying that the mean success rate is the highest across all tolerance levels. The multi-channel architecture recovers the degradation (dimming) factor for all channels with a success rate of at least $91\%$ for a relative tolerance level of $20\%$ and a mean success rate of $\sim94\%$. It is also evident that this model outperforms the baseline and the single-channel model for all levels of relative tolerances.  For any given level of tolerance, the mean across all channels increased significantly. For example, with absolute tolerance of $0.05$, the mean increase from $78\%$ to $85\%$, even changing its color classification. In addition, the success rate is consistently the worst for $335$~\AA~and $211$~\AA~channels across all tolerances, whereas the performance of the $131$~\AA~channel is the best.

Looking at specific channels, we can see that $304$~\AA~ does consistently well through all the models with not much variation, which wasn't expected. Now observing $171$~\AA, it does well in the baseline and in the multi-channel model, but surprisingly it has its maximum performance in the single-channel model, through all tolerances, and a remarkable $94\%$ success rate with a tolerance of $0.05$. In opposite to $171$~\AA, channels $211$~\AA~and $335$~\AA~ have a poor performance in the baseline and the single-channel models, and they have a significant improvement in the multi-channel model as expected and hypothesized by this paper.

\begin{figure}
    \centering
        \includegraphics[width=0.5\textwidth]{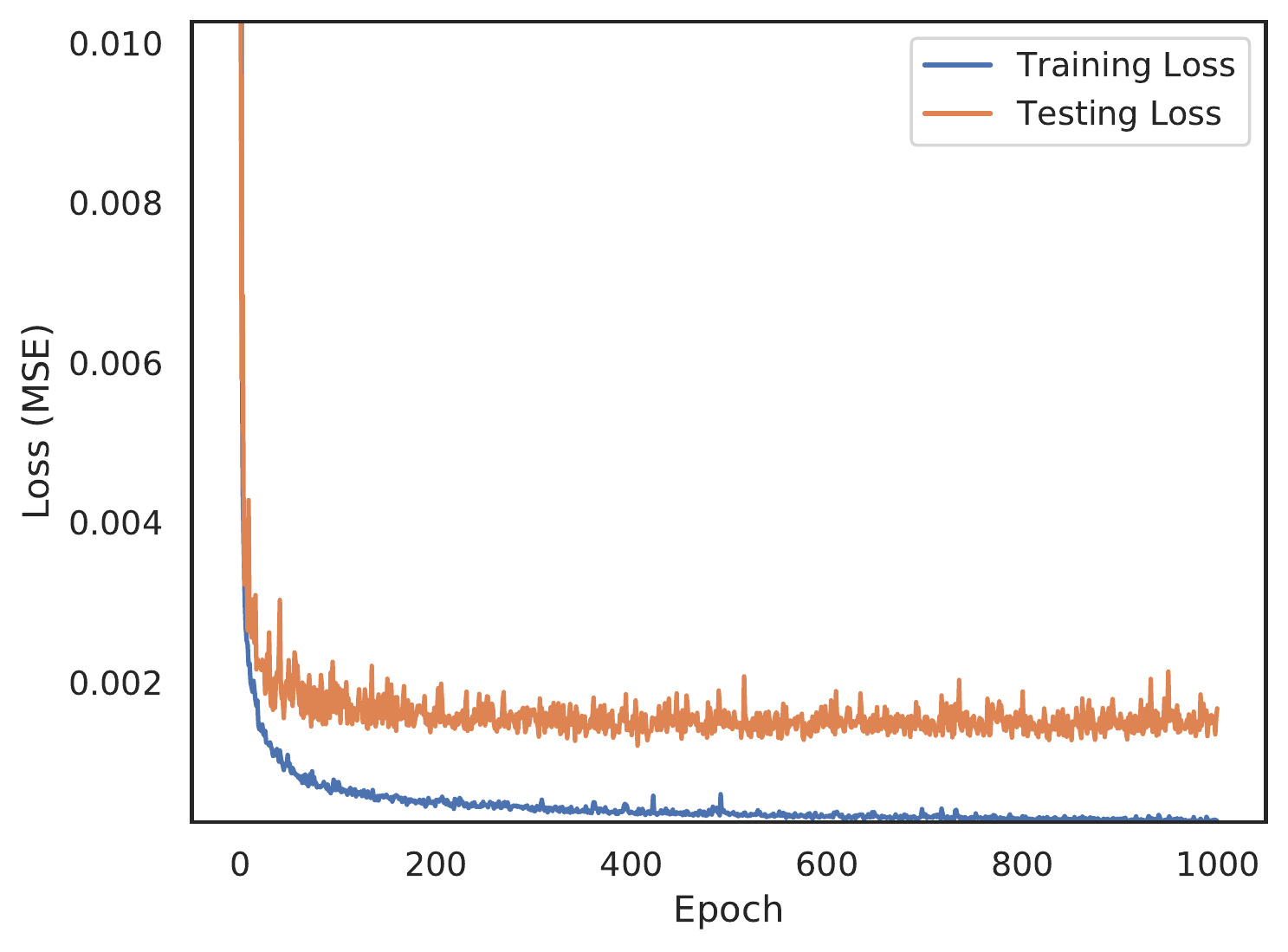}
        \caption{Graphic of the evolution of training and testing MSE loss through the epochs.}
        \label{fig:training_curve}
\end{figure}

Observing the Fig.\ref{fig:training_curve}, we can see the training and test MSE loss curve evolving by epoch. Based on the results from Table \ref{tab:autocalibrate_final_results} and comparing the training and test loss curves in Fig. \ref{fig:training_curve} we can see the model does not heavily overfit in the range of epochs utilized, and it presents stable generalization performance on test results. We stopped the training before epoch 1000, seeing only marginal improvements achieved in the test set over many epochs.

Overall the result shows higher success rates for the CNN models, particularly for the multi-channel model, which was predicted by the toy problem, and for higher tolerances.

\subsection{Modelling Channel Degradation over Time}
    \label{sec:degradation}
    
In this section, we discuss the results obtained when comparing the AIA degradation curves $V8$ and $V9$, with both single-channel and multi-channel CNN models. This process was performed using a dataset equivalent to the SDOML but with no correction for degradation and data period from $2010$ to $2020$. This tests both models for real degradation suffered by AIA from $2010$ to $2020$.

\begin{figure*}
	\centering
		\centering
        \includegraphics[width=0.98\textwidth]{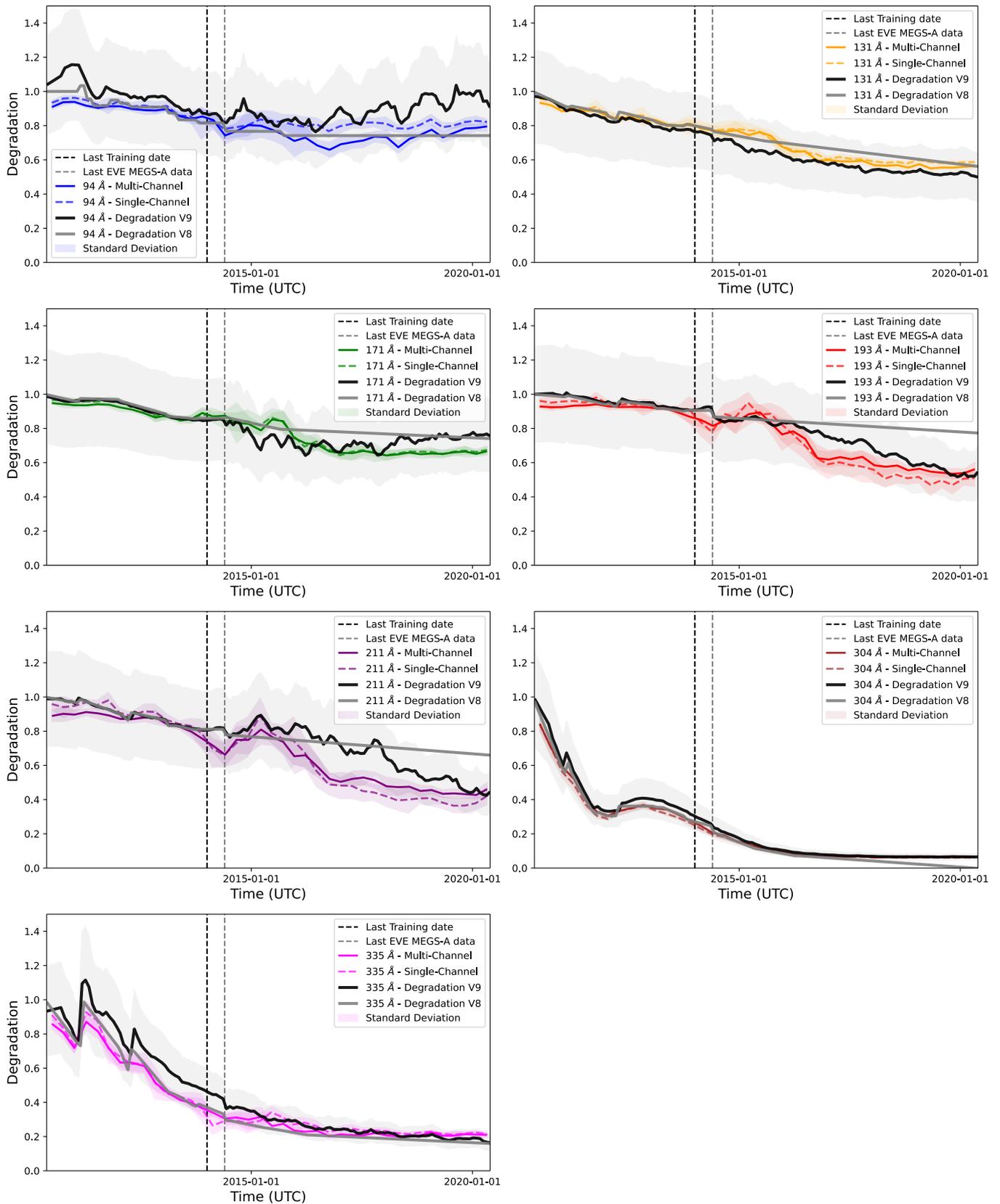} 
        \caption{Channels degradation over Time. From top to bottom: Channel $94$~\AA~ (blue) and $131$~\AA~(yellow), $171$\AA~(green) and $193$~\AA~(red), $211$~\AA~(purple) and $304$~\AA~(brown) and $335$~\AA~(magenta). The solid black (gray) curve is the degradation profile of AIA calibration release $V9$ ($V8$). The gray shaded area correspond to the $25\%$ error of the degradation curve $V9$. The colorful shaded areas are the standard deviation of the CNN models. The vertical black dashed line is the last available observation from EVE MEGS-A data and the vertical gray dashed line is the last training date.}
        \label{fig:degradation_curve}
\end{figure*}

Figure~\ref{fig:degradation_curve} presents the results of our analysis for all the seven AIA EUV channels. In each panel, we show four quantities: the degradation curve $V9$ (solid black line), the degradation curve $V8$ (solid gray line), predicted degradation from the single-channel model (dashed colorful line), and multi-channel model (solid colorful line). The shaded gray band depicts the region covering $25\%$ variation (error) associated with the $V9$ degradation curve, and the colorful shaded areas are the standard deviation of the single- and multi-channel models. The dashed vertical line coincides with the day (25 May 2014), the last day of EVE MEGS-A instrument data. It is important to note that MEGS-A was earlier used for the sounding rocket calibration purposes, the loss of which caused both the $V8$ and $V9$ degradation curves to become noisier in the future. \citet{Szenicereaaw6548} used deep learning to facilitate a virtual replacement for MEGS-A.

Observing the different panels of Fig.~\ref{fig:degradation_curve}, we can see that even though we trained both the single and multi-channel models with the SDOML dataset that was produced and corrected using the $V8$ degradation curve, both CNN models predict the degradation curves for each channel quite accurately over time, except for $94$~\AA\ and $211$~\AA\ channel. However, the deviations of the predicted values for these two channels fall well within the $25\%$ variation of the $V9$  calibration curve. In fact, the CNN predictions have even better agreement with $V9$ than the $V8$ calibration for most of the channels. That hints at the conclusion that the CNN is picking up on some actual information that is perhaps even more responsive to degradation than FISM. The latest degradation curve ($V9$) was updated recently in July $2020$, and the change from $V8$ to $V9$ might have easily caused an impact while training the models. Moreover, the more significant deviation of $94$~\AA\ channel in the early stages of the mission is due to the fact we limited our degradation factor to be less than one. 

From the predicted calibration curves computed from the single- and multi-channel models, we see that they have a significant overlap throughout the entire period of observation. The single-channel model predictions, however, have a more significant variation for channels $211$~\AA, $193$~\AA~ and $171$~\AA. For a systematic evaluation and a comparison among the results of the two models across channels, we calculated some goodness of fit metrics, and the results are shown in Table~\ref{tab:quantities_degradation}.

\begin{table}[h]
  \centering
  \caption{Goodness of fit metrics for single-channel and multi-channel models with reference to the $V9$ degradation curve. The first metric is the Two-Sample Kolmogorov-Smirnov Test (KS), and the second metric is the Fast Dynamic Time Warping.}
  \label{tab:quantities_degradation}
  \centering
  \begin{tabular}{ccccc}
    \toprule
    \multirow{2}{*}{Channel} & \multicolumn{2}{c}{{\centering Single-Channel}} &  \multicolumn{2}{c}{{\centering Multi-Channel}}  \\
    \cmidrule(lr){2-3}\cmidrule(lr){4-5}
     & KS & DTW &  KS & DTW\\
     \midrule
     94~\AA  &  0.485 & 7.120 &  0.568 & 9.624\\
     131~\AA &  0.346 & 2.711 &  0.275 & 1.624\\
     171~\AA &  0.298 & 3.074 &  0.329 & 3.549\\
     193~\AA &  0.211 & 1.829 &  0.244 & 2.080\\
     211~\AA &  0.305 & 2.850 &  0.242 & 2.807\\
     304~\AA &  0.282 & 1.412 &  0.100 & 1.311\\
     335~\AA &  0.212 & 2.539 &  0.141 & 2.839\\
     \bottomrule
  \end{tabular}
\end{table}

Table \ref{tab:quantities_degradation} contains two different metrics for evaluating the goodness of fit of each CNN model with the $V9$ degradation curve. The first is the Two-Sample Kolmogorov--Smirnov Test (KS), which determines whether two samples come from the same distribution \citep{two_ks}, and the null hypothesis assumes that the two distributions are identical. The KS test has the advantage that the distribution of statistics does not depend on the cumulative distribution function being tested. The second metric is the Fast Dynamic Time Warping \citep[DTW, ][]{fastDTW}, which measures the similarity between two temporal sequences that may not be of the same length. This last one is important since statistical methods can be too sensitive when comparing both time series. DTW has distance between the series as an output, and as a reference the DTW for the different EUV channels between the $V8$ and $V9$ degradation curves are: $94$~\AA-$72.17$, $131$~\AA-~$13.03$, $171$~\AA-$9.82$, $193$~\AA-$30.05$, $211$~\AA-$16.86$, $304$~\AA-$7.02$ and $335$~\AA-$5.69$.

Similar to Fig.~\ref{fig:degradation_curve} we find in Table~\ref{tab:quantities_degradation}, that the predictions from both the single-channel and multi-channel models overlap significantly both in terms of the metric and the time evolution. Except for the $94$~\AA~channel, all others have very close metric values, well within a given level of tolerance. A low value of the KS test metric suggests that the predictions have a similar distribution as the observed $V9$ calibration curve, which also indicates the robustness of our CNN architecture. KS test agrees well with DTW, where the values obtained are smaller than the reference values (as indicated earlier) between the $V8$ and the $V9$ calibration curves. Overall, the metric analysis for the goodness of fit between the predictions and the actual calibration curve ($V9$) shows that the CNN models perform remarkably well in predicting the degradation curves despite being trained only on the first three years of the observations.  

\subsection{Feature Maps}
    \label{sec:feature-maps}
    
\begin{figure} [h!]
  \centering
  \includegraphics[width=0.8\linewidth]{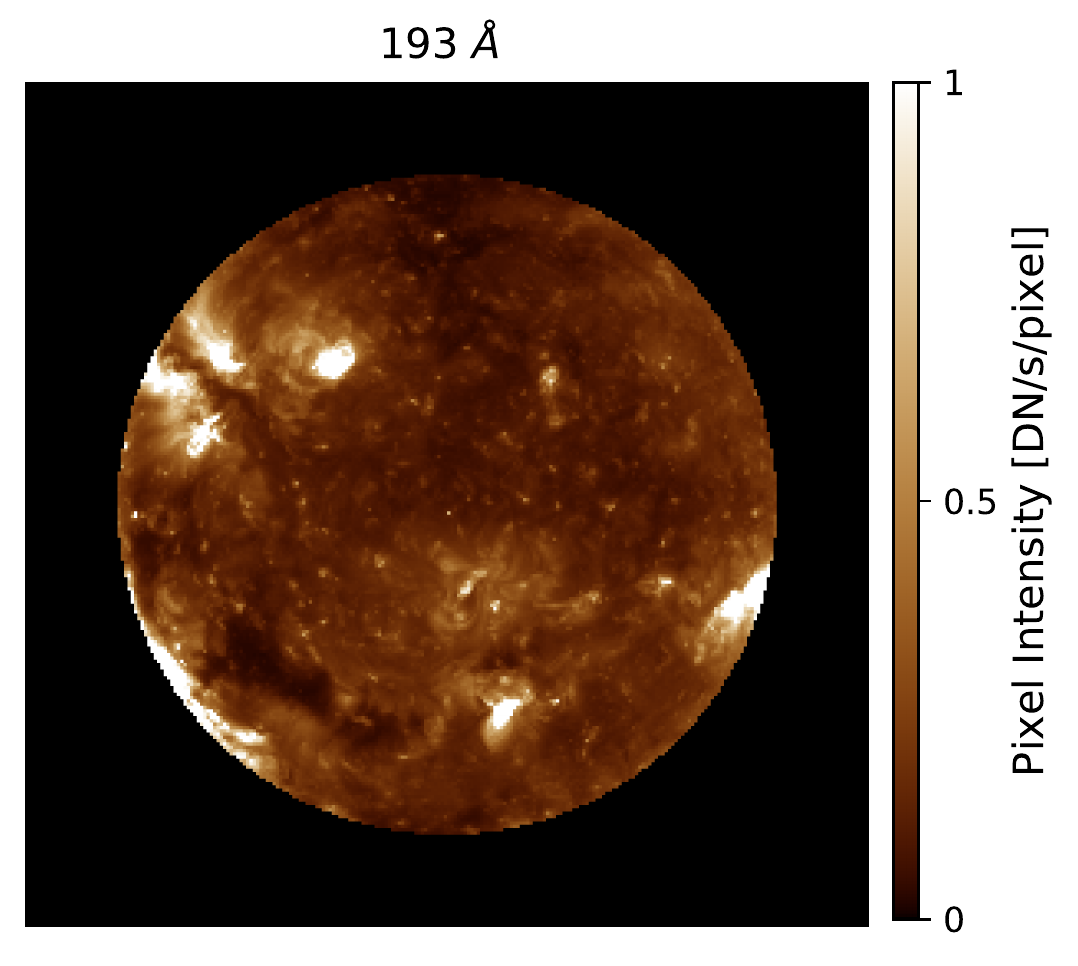}
  \includegraphics[width=\linewidth]{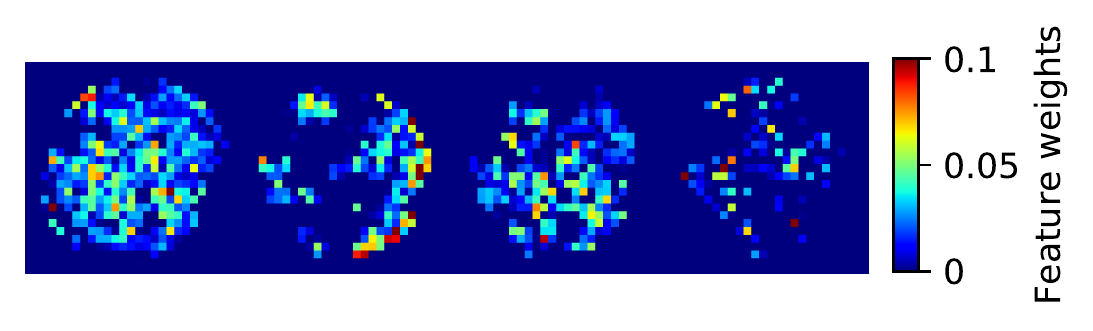}
  \caption{Feature maps obtained from the last layer of CNN of our model. The top row shows a sample input in AIA 193~\AA~ channel, and the bottom row shows four representative feature maps out of one hundred and twenty eight different feature maps from the final convolutional layer of the multi-channel NN model.}
  \label{fig:autocalibrate_activation_viz}
\end{figure}

As mentioned in Sect.~\ref{section:model-benchmark-understanding}, the feature maps are the result of applying the filters to an input image. That is, at each layer, the feature map is the output of that layer. In Fig.~\ref{fig:autocalibrate_activation_viz} we present such maps obtained from the output of the last convolutional layer of our CNN. The top row shows a reference input image observed at $193$~\AA\ used in this analysis, with its intensity scaled between $0-1$ pixel units, and the bottom row shows $4$ representative feature maps (out of a total of $128$) with their corresponding weights. These maps are obtained after the final convolutional layer of the multi-channel model, and it represents the result of combining all seven EUV channels as input. The predicted $\alpha$ dimming factors from the model are given by the sigmoid activation function applied to a linear combination of these features. Such mapping allows us to see that the network actually learned to identify the different features of such full-disk solar images such as the limb, the quiet Sun features, and the ARs. The reason for visualizing a feature map for specific AIA images is to gain an understanding of what features a model detects are ultimately useful in recovering the degradation or the dimming factors.

\section{Concluding remarks}
\label{section:summary}
This paper reports a novel ML-based approach to auto-calibration and advances our comprehension of the cross-channel relationship among different EUV channels by introducing a robust novel method to correct for the EUV instrument time degradation. We began with formulating the problem and setting up a toy model to test our hypothesis. We then established two CNN architectures that consider multiple wavelengths as input to auto-correct for on-orbit degradation of the AIA instrument onboard SDO. We trained the models using the SDOML dataset and further augmented the training set by randomly degrading images at each epoch. This approach made sure that the CNN model generalizes well to data not seen during the training, and we also developed a non-ML baseline to test and to compare its performance with the CNN models. With the best trained CNN models, we reconstructed the AIA multi-channel degradation curves of 2010-2020 and compared them with the sounding-rocket based degradation curves $V8$ and $V9$.

Our results indicate that the CNN models significantly outperform the non-ML baseline model ($85\%$ vs. $51\%$ in terms of the success rate metric), for a tolerance level of $0.05$. In addition, the multi-channel CNN also outperforms the single-channel CNN with a $78\%$ success rate with an absolute $0.05$ threshold. This result is consistent with the expectation that correlations between structures in different channels, and size (morphology) of structures, and brightness can be used to compensate for the degradation. To further understand the correlation between different channels, we used the concept of feature maps to shed light over this aspect and see how the filters of the CNNs were being activated. We did see that the CNNs learned representations that make use of the different features within solar images, but further work needs to be done in this aspect to establish a more detailed interpretation.

We also found that the CNN models reproduce the most recent sounding-rocket based degradation curves ($V8$ and $V9$) very closely and within their uncertainty levels. This is particularly promising, given that no time information has been used in training the models. For some specific channels, like $335$~\AA, the model is reproducing the $V8$ curve instead of $V9$ since the SDOML was corrected using the former. The single-channel model could perform as well as the multi-channel model even though the multi-channel presented a more robust performance when evaluated on the basis of their success rates.

Lastly, this paper presents a unique possibility of auto-calibrating deep space instruments such as the ones onboard the STEREO spacecraft, and the recently launched remote sensing instrument called \textit{Extreme Ultraviolet Imager} \citep{2020A&A...642A...8R}, aboard the Solar Orbiter satellite \citep{2020A&A...642A...1M}, that are too far away from the Earth to be calibrated using a traditional method such as sounding-rockets. The auto-calibration model could be trained using the first months of data from the mission, assuming the instrument is calibrated at the beginning of the mission. The data volume could be an issue, and different types of data augmentation could be used to overcome this problem, such as synthetic degradation and image rotation. We further envision that the technique presented here may also be adapted to imaging instruments or spectrographs operating at other wavelengths (e.g., hyperspectral Earth-oriented imagers) observed from different space-based instruments like \textit{IRIS} \citep[][]{2014SoPh..289.2733D}. 

\begin{acknowledgements}

{This project was partially conducted during the 2019 Frontier Development Lab (FDL) program, a co-operative agreement between NASA and the SETI Institute. We wish to thank IBM for providing computing power through access to the Accelerated Computing Cloud, as well as NASA, Google Cloud and Lockheed Martin for supporting this project. L.F.G.S was supported by the National Science Foundation under Grant No. AGS-1433086. M.C.M.C. and M.J. acknowledge support from NASA’s SDO/AIA (NNG04EA00C) contract to the LMSAL. S.B.  acknowledges the support from the Research Council of Norway, project number 250810, and through its Centers of Excellence scheme, project number 262622. This project was also partially performed with funding from Google Cloud Platform research credits program. We thank the NASA’s Living With a Star Program, which SDO is part of, with AIA, and HMI instruments on-board. CHIANTI is a collaborative project involving George Mason University, the University of Michigan (USA), University of Cambridge (UK) and NASA Goddard Space Flight Center (USA). A.G.B. is supported by EPSRC/MURI grant EP/N019474/1 and by Lawrence Berkeley National Lab. The authors thank the anonymous referee for the comments.\\

Software: We acknowledge for CUDA processing cuDNN \citep{cudnn}, for data analysis and processing we used Sunpy \citep[][]{Sunpy2020}, Numpy \citep{numpy}, Pandas \citep{pandas}, SciPy \citep{scipy}, scikit-image \citep{scikit-image} and scikit-learn\citep{scikit-learn}. Finally all plots were done using Matplotlib \citep{matplotlib} and Astropy \citep{astropy:2018}}.
\end{acknowledgements}

\bibliographystyle{aa}
\bibliography{main}

\begin{appendix}
\section{Scaling Units for each AIA channel}
\label{section:appendix_average}

\begin{table}[h!]
  \centering
  \caption{Table of the scaling units of  AIA channels.}
  \label{tab:average_channels}
  \begin{tabular}{cc}
    \toprule
     AIA channel (\AA) &  Scaling unit [DN/s/pixel] \\
     \midrule
       94 &   10  \\
      131 &   80  \\
      171 & 2000  \\
      193 & 3000  \\
      211 & 1000  \\
      304 &  500  \\
      335 &   80  \\
      \bottomrule
  \end{tabular}
\end{table}

\section{Detailed model Architectures}
\label{section:appendix_archtectures}

Table \ref{tab:single-channel-arch} and \ref{tab:multi-channel-arch} presents the more detailed information of the CNN architecture. From left to right, they show the layer number, the type of layer, the output shape of each layer, and the learnable parameters each layer has.  During the training process, the model works to learn and optimize the weights and biases in a neural network. These weights and biases are indeed the learnable parameters. In fact, any parameters within our model that are learned/updated during the training process is a learnable parameter.

\begin{table}[h!]
  \centering
  \caption{The detailed single channel architecture. From left to right columns, we have the layer number, layer type, output shape, and the total number of learnable parameters of the layer. The architecture is composed of two blocks of a convolutional layer, ReLU activation function and max pooling layer, followed by a fully connected (FC) layer and a final sigmoid activation function.}
  \label{tab:single-channel-arch}
  \begin{tabular}{|c|c|c|c|}
    \hline
    \# & Layer &  Output Shape & Learnable Parameters \\
     \hline
1&           Input    &  [1, 256, 256]    &         0 \\
2&     Convolution    & [64, 254, 254]    &       640 \\
3&            ReLU    & [64, 254, 254]    &         0 \\
4&      MaxPooling    &   [64, 84, 84]    &         0 \\
5&     Convolution    &  [128, 82, 82]    &    73,856 \\
6&            ReLU    &  [128, 82, 82]    &         0 \\
7&      MaxPooling    &  [128, 27, 27]    &         0 \\
8& Fully Connected    &            [1]    &    93,313 \\
9&         Sigmoid    &            [1]    &         2 \\  
            \hline
\multicolumn{3}{|c|}{Total Learnable Parameters} &   167,809 \\

    \hline
  \end{tabular}
\end{table}

\begin{table}[h!]
  \centering
  \caption{The detailed multi-channel architecture. From left to right columns, we have the layer number, layer type, output shape, and the total number of learnable parameters of the layer. The architecture is composed of two blocks of a convolutional layer, ReLU activation function and max pooling layer, followed by a fully connected (FC) layer and a final sigmoid activation function.}  
  \label{tab:multi-channel-arch}
  \begin{tabular}{|c|c|c|c|}
    \hline
     \# & Layer &  Output Shape & Learnable Parameters \\
     \hline
1&           Input    &  [7, 256, 256]    &         0 \\
2&     Convolution    & [64, 254, 254]    &     4,096 \\
3&            ReLU    & [64, 254, 254]    &         0 \\
4&      MaxPooling    &   [64, 84, 84]    &         0 \\
5&     Convolution    &  [128, 82, 82]    &    73,856 \\
6&            ReLU    &  [128, 82, 82]    &         0 \\
7&      MaxPooling    &  [128, 27, 27]    &         0 \\
8& Fully Connected    &            [7]    &   653,191 \\
9&         Sigmoid    &            [7]    &        56 \\
            \hline
\multicolumn{3}{|c|}{Total Learnable Parameters} &   731,143 \\

    \hline
  \end{tabular}
\end{table}

Learnable parameters are calculated differently for each layer. For convolutional layers we use equation \ref{appendix:equation_conv}

\begin{equation}
     LP_{Convolution} = ((H \times W \times f) + 1) \times k)
     \label{appendix:equation_conv}
\end{equation}

\noindent where $LP_{Convolution}$ are the learnable parameters of the convolution layer, $H$ is the height of the input, $W$ is the width, $f$ is the number of filters from the previous layer, $1$ is the bias, and $k$ is the number of filters in the convolution. For the fully connected layers and sigmoids, we use the equation \ref{appendix:equation_fc}

\begin{equation}
    LP_{Connected} = ((C \times P) + 1 \times c)
    \label{appendix:equation_fc}
\end{equation}

\noindent where $LP_{Connected}$ are the learnable parameters of the fully connected layer $C$ is the number of current layer neurons, $P$ is the number of previous layers neurons, and $1$ is the bias. ReLU and max-pooling layers have zero learnable parameters since they do not have weights to be updated as the neural network is trained.
\end{appendix}
\end{document}